\documentclass[11pt]{article}

\usepackage{graphicx}
\begin{document}
\oddsidemargin .03in
\evensidemargin 0 true pt
\topmargin -.4in

%Abbreviations %
%***************%

\def\ra{{\rightarrow}}
\def\a{{\alpha}}
\def\b{{\beta}}
\def\l{{\lambda}}
\def\eps{{\epsilon}}
\def\T{{\Theta}}
\def\t{{\theta}}
\def\co{{\cal O}}
\def\car{{\cal R}}
\def\caf{{\cal F}}
\def\cs{{\Theta_S}}
\def\pr{{\partial}}
\def\tri{{\triangle}}
\def\na{{\nabla }}
\def\S{{\Sigma}}
\def\s{{\sigma}}
\def\sp{\vspace{.1in}}
\def\hs{\hspace{.25in}}

\newcommand{\be}{\begin{equation}} \newcommand{\ee}{\end{equation}}
\newcommand{\bea}{\begin{eqnarray}}\newcommand{\eea}
{\end{eqnarray}}

%********************************************************************%

\begin{titlepage}
\topmargin= -.2in
\textheight 9.6in

\begin{center}
\baselineskip= 15 truept
\vspace{.3in}

\centerline{\Large\bf Emergent ${\mathbf D}$-instanton as a source of Dark Energy}
%\centerline{\Large\bf Can a vacuum  pair of $(3{\bar 3})$-brane world provide a clue?}

\vspace{.3in}
\noindent
{{\bf $^{*}$Deobrat Singh$^{a,b,}$}\footnote{deobratsingh10@gmail.com, deobratsingh10@mitsgwalior.in, Mob.: +91-9891801527} {\bf and Supriya  Kar$^{a}$}}

\vspace{.15in}

\noindent

\noindent
{{\large $^{a}$Department of Physics \& Astrophysics}\\
{\large University of Delhi, New Delhi 110 007, India}}\\
\noindent

\noindent
{{\large $^{b}$Department of Applied Science}\\
{\large Madhav Institute of Technology and Science, Gwalior, India}}

\vspace{.3in}
\noindent
{{\bf Short running title:} Emergent ${D}$-instanton as a source of Dark Energy}

\vspace{.2in}

%{\today}
\thispagestyle{empty}

\vspace{.4in}
\begin{abstract}
We revisit a non-perturbative formulation leading to a vacuum created gravitational pair of $(3{\bar 3})$-brane by a Poincare dual higher form $U(1)$ gauge theory on a $D_4$-brane. In particular, the analysis has revealed a dynamical geometric torsion ${\cal H}_3$ for an onshell Neveu-Schwarz (NS) form on a fat $4$-brane. We argue that a $D$-instanton can be a viable candidate to incorporate the quintessence correction hidden to an emergent $(3+1)$-dimensional brane universe. It is shown that a dynamical non-perturbative correction may be realized with an axionic scalar QFT on an emergent anti $3$-brane within a gravitational pair. The theoretical tool provokes thought to believe for an extra instantaneous dimension transverse to our classical brane-universe in an emergent scenario. Interestingly a $D$-instanton correction, sourced by an axion on an anti $3$-brane, may serve as a potential candidate to explain the accelerated rate of expansion of our $3$-brane universe and may provide a clue to the origin of dark energy.

\vspace{.2in}

\baselineskip=14 truept

\vspace{.12in}

\vspace{1in}

\noindent

\noindent

\end{abstract}
\end{center}

\vspace{.2in}

\baselineskip= 16 truept

\vspace{1in}

\end{titlepage}

\baselineskip= 15 truept

%%%%%%%%%%%%%%%%%%%%%%%%%%%%%%%%%%%%%%%%%%%%%%%%%%%%%%%%%%%%%
\section{Introduction}
%%%%%%%%%%%%%%%%%%%%%%%%%%%%%%%%%%%%%%%%%%%%%%%%%%%%%%%%%%%%%

\vspace{-.1in}
\noindent
The origin of the current accelerated expansion of the Universe is one of the biggest problem in modern
physics \cite{{perlmutter},{supernova}} (see \cite{{spergel},{sdss},{sdss1},{ade}} for more). To explain accelerated expansion of the Universe, one might introduce an unknown energy source, i.e., dark energy \cite{sami}. The easiest way to do this is the cosmological constant ($\Lambda$), and
introducing another invisible component, dark matter. The model with these components is known as the $\Lambda$CDM model or standard model of cosmology where CDM stands for Cold Dark Matter. this model is well compatible with observational findings. However, fine-tuning problem of the cosmological constant is a big suffering for the $\Lambda$CDM model. The mystery
of the accelerated expansion of the Universe thus motivates us to explore the possibilities of modification of general relativity on cosmological scales \cite{clifton}. Among a number of recent developments, the emergent gravity \cite{yang} discussed in the recent past could be very interesting.

\sp
\noindent
The General Theory of Relativity (GTR) is a classical field description of an intrinsic metric on a Riemannian manifold. Black holes are well defined theoretical perspectives consistently described by the non-linear differential equations of motion of the metric field. Importantly the contracted Bianchi identities ensure the conservation of energy-momentum-stress tensor and is believed to be governed by a non-linear field. Furthermore the observed cosmology, underlying an accelerated rate of expansion of universe, presumably demands an entanglement of quantum mechanics with the GTR. However the perspective of linear operators or superposition principle in quantum mechanics prohibits a consistent quantum description of the GTR. In addition the GTR is an interacting metric tensor field theory and hence perturbative quantum corrections would further bring-in instability and hence the theory breaks down.

\sp
\noindent
In the context superstring theories are viable candidates to describe quantum gravity consistently. Interestingly the stringy versions of these macroscopic black holes have been obtained in a low energy limit of a string effective action in the past \cite{candelasHS,freed,callan,garfinkle,giddings-strominger}. In fact, a nontrivial background metric on a string world-sheet is known to incorporate vertex operators in a path integral and hence naturally introduces a string coupling at each interaction vertex. A weak coupling consistently incorporates perturbative geometric (higher genus) corrections to the string world-sheet at tree level and leads to a theory of quantum gravity. 

\sp
\noindent
On the other hand the quantum effects to Einstein gravity can also be explored addressed non-perturbatively using the strong-weak coupling duality in ten dimensional superstring theories \cite{ashoke-sen-ijmpa}. A non-perturbative quantum effect is believed to be sourced by compactified extra space dimension(s) to various stringy vacua. In particular, the type IIA superstring theory in a strong coupling limit is known to incorporate an extra spatial dimension on $S^1$ and has been identified with the non-perturbative $M$-theory in eleven dimensions. Generically $M$-theory has been shown to be identified with the stringy vacua in various dimensions \cite{witten-M}. In a low energy limit $M$-theory is known to describe an eleven dimensional supergravity. 

%%%%%%%%%%%%%%%%%%%%%%%%%%%%%%%%%%%%%%%%%%%%%%%%%%%%%%%%%%%%%%%%%%%
\section{Pair creation of $(3{\bar 3})$-brane universe at Big Bang}
%%%%%%%%%%%%%%%%%%%%%%%%%%%%%%%%%%%%%%%%%%%%%%%%%%%%%%%%%%%%%%%%%%%
We begin by briefly revisiting the Schwinger pair production mechanism \cite{schwinger} established in quantum field theory (QFT). The novel idea elegantly describes non-perturbative particle production from vacuum. A positron and electron pair is produced at a QFT vacuum by an electromagnetic quanta, $i.e.$ photon. The quantum vacuum is unstable under the influence of an external electric field, as the virtual electron-positron dipole pairs can gain energy from the external field. If the field is sufficiently strong, these virtual particles can gain the threshold pair creation energy $2mc^2$ and become real electron-positron pairs.The pair production mechanism is an emergent non-perturbative phenomenon as the virtual particles tunnel out of the Dirac sea. It is believed to be instrumental to describe diverse quantum phenomenon in gravitation and cosmology, in particular for the phenomena of Hawking radiation near a black hole, and Unruh radiation of accelerating mirrors. It is also closely related to the dynamical Casimir effect of atomic physics. In fact the tool was explored to explain the Hawking radiation phenomenon \cite{hawking}, Virtual particle pairs are constantly being created near the horizon of the black hole, as they are everywhere.  Normally, they are created as a particle-antiparticle pair and they quickly annihilate each other.  But near the horizon of a black hole, it's possible for one to fall in before the annihilation can happen, in which case the other one escapes as Hawking radiation. Far from the black hole, the gravitational effects can be weak enough for calculations to be reliably performed in the framework of quantum field theory in curved spacetime. Hawking showed that quantum effects allow black holes to emit exact black body radiation. Furthermore the idea was applied to the open strings pair production by Bachas and Poratti \cite{bachas-porrati}. Majumdar and Davis have exploited the underlying mechanism to investigate the production of a higher dimensional pair of $(D{\bar D})_9$ at the cosmological horizon \cite{majumdar-davis} using Kibble-Zurek mechanism (KZM) \cite{kibble}. KZM basically describes the non-equilibrium dynamics and the formation of topological defects in a system which is driven through a continuous phase transition at finite rate. The KZM generally applies to spontaneous symmetry breaking scenarios where a global symmetry is broken. KZM has applications not only in classical  but also in quantum phase transitions and even in optics.

\sp
\noindent
People in String theory, are continuously working to remove the shortcomings of Standard model of Cosmology including inflation and to provide a better explanation of early universe cosmology \cite{veneziano}.
An extra spatial dimension compactified on $S^1/Z2$ has been argued to unfold in the ten dimensional type IIA superstring theory when the string coupling becomes large. As a result the strong coupling string theory has been conjectured to describe an eleven dimensional non-perturbative $M$-theory \cite{horava-witten}.

\sp
\noindent 
Interestingly the emergent curvature on a pair of brane/anti-brane was shown to be described by a geometric torsion in a string-brane model \cite{abhishek-JHEP,abhishek-PRD,abhishek-NPB-P}. The model may be viewed as an attempt to explore the non-perturbative domain in quantum gravity underlying a string theoretic setup. Importantly the torsion geometry underlying a $(4+1)$-dimensional world-volume gauge theory was generalized to $(5+1)$-dimensional CFT to describe a gravitational pair of de Sitter $(4{\bar 4})$-brane universe \cite{{richa-IJMPD},{deobrat}}. 

%%%%%%%%%%%%%%%%%%%%%%%%%%%%%%%%%%%%%%%%%%%%%%%%%%%
\subsection{Dynamical NS field in string-brane setup}
%%%%%%%%%%%%%%%%%%%%%%%%%%%%%%%%%%%%%%%%%%%%%%%%%%%
The necessary step to formalize the mathematical implementation of torsion
terms in braneworld scenarios, has already been presented in \cite{daSilva} and authors have tried to apply their results in cosmological problems. In another paper by authors, it has been argued that the torsionless brane universe may naturally be substituted by a more congenial braneworld scenario that contains torsion, and may be useful to a more precise description of physical theories \cite{daSilva1}. The work presents the torsion effects in braneworld scenarios in a very consistent and appealing way. The warped extra-dimensional formalism points to the presence of new interactions, of significant phenomenological
importance, between the Kaluza-Klein modes of the dilaton and the KR field. In the recent past a geometric torsion ${\cal H}_3$ has been constructed to address a non-perturbative \cite{polchinski} quantum correction to the low energy string vacuum \cite{abhishek-JHEP,abhishek-PRD,abhishek-NPB-P}. It was shown that non-BPS brane may well be described by a non-perturbative phenomenon leading to a vacuum created gravitational pair of $(3{\bar 3})$-brane by the Kalb-Ramond (K-R or KR) two form quanta on a $D_4$-brane at the cosmological horizon of a background de Sitter brane geometry. The absorption of the K-R field (quantum) in the five dimensional $U(1)$ gauge theory was shown to modify the covariant derivative, which in turn describes an emergent torsion curvature on a pair. Alternately, K-R field local degrees have been transformed to project a dynamical Neveu-Schwarz (NS) field on a vacuum created non-BPS pair of $(D{\bar D})_3$-brane which presumably defines our early universe. The momentum conservation ensures that the vacuum created gravitational $3$-brane and ${\bar 3}$-brane cannot annihilate each other and defines a stable pair of $(3+1)$-dimensional brane/anti-brane universe. In fact, the cosmological creation of a pair of non-BPS $(D{\bar D})_9$-brane in a different context was discussed by Majumdar and Davis \cite{majumdar-davis}. Their work was primarily motivated to address the strong coupling regime in type IIA superstring theory leading to $M$-theory. It is worth mentioning that a pair of brane and anti-brane is known to break the supersymmetry and hence is described by a non-BPS brane. Past analysis has revealed that some of the non-BPS states are stable in superstring theories \cite{Ashoke-Sen}. It was shown that the (open string) tachyon condenses on a non-BPS brane to describe a stable pair of $(D{\bar D})$-brane \cite{Ashoke-Sen-P}.

\sp
\noindent
In the context the vacuum refers to an open string background black hole on a $D_4$-brane which is known to be sourced by a constant NS field \cite{seiberg-witten} in the string bulk. Since K-R potential is a two form potential, The KR field quanta is believed to vacuum create a pair of $(1{\bar 1})$-brane instead of a charged pair of particle and anti-particle which are created by one form field quanta. Their stability favors an increase in the number of their spatial dimensions. Finally, they describe a space filling gravitational pair of $(3{\bar 3})$-brane underlying the $(4+1)$-dimensional world-volume gauge theory. The KR field dynamics is given by
\be
S=- {1\over{12C_2^2}}\int d^5x\ {\sqrt{-{\tilde G}^{({\rm NS})}}}\ H_{\mu\nu\lambda}H^{\mu\nu\lambda}\ ,\qquad {\rm where}\quad C_2^2=(8\pi^3g_s){\alpha'}^{3/2}\ .\label{gauge-2}
\ee
The significance of the constant NS field in addition to the local KR field have been addressed in an effective curvature formalism on a $D_4$-brane \cite{abhishek-JHEP,abhishek-PRD,abhishek-NPB-P}. Importantly $H_3$ has been identified with a connection which in turn modifies the covariant derivative to ${\cal D}_{\mu}$ on a $D_4$-brane. The mathematical construction implies an absorption of KR field quanta and hence the ${\cal D}_{\mu}$ describes a ``fat'' brane or a string-brane model. In other words the model evolves with a dynamical NS field at the expense of the KR field on a $D_4$-brane. As a result the emergent pair of brane-universe is governed by a geometrical torsion: 
\bea
{\cal H}_{\mu\nu\lambda}&=&{\cal D}_{\mu}B^{\rm NS}_{\nu\lambda}+{\rm cyclic}=\left ( H_{\mu\nu\alpha}{B^{({\rm NS})\alpha}}_{\lambda} + \rm{cyclic}\right ) 
+ H_{\mu\nu\beta} {B^{({\rm NS})\beta}}_{\alpha} {B^{({\rm NS})\alpha}}_{\lambda} +\ \dots\ ,\nonumber\\
{\rm where}&&{\cal D}_{\lambda}B^{\rm NS}_{\mu\nu}= {1\over2}{H_{\lambda\mu}{}}^{\rho}B^{\rm NS}_{\rho\nu} - {1\over2} {H_{\lambda\nu}{}}^{\rho}B^{\rm NS}_{\rho\mu}\quad {\rm and}\;\nabla_{\lambda}B^{{\rm NS}}_{\mu\nu}=0\ .\label{gauge-22}
\eea
It shows that the geometric torsion underlies a fat brane which breaks the supersymmetry. Thus it describes a non-BPS pair of emergent gravitational $(3{\bar 3})$-brane. The presence of string bulk due to the dynamical NS field ensures the closed string modes coupling to the otherwise BPS brane. Importantly, the geometric torsion in the string-brane model is exact and hence is non-perturbative. In addition, the $U(1)$ gauge invariance of a geometric torsion under a two form transformation has been realized in an effective space-time curvature formalism. A dynamical space-time without a torison in $4D$ has been argued to began at a Big Bang singularity defined at a cosmological horizon. It is noteworthy to remark that the space-time curvature sourced by a geometric torsion in $5D$ naturally incorporates an intrinsic angular momentum on a vacuum created pair of $(3{\bar 3})$-brane. The world-volumes of a brane and an anti-brane are separated by an extra fifth dimension transverse to each other. It may imply that an observer in a $3$-brane universe is inaccessible to an anti $3$-brane universe or vice-versa. Nevertheless, the hidden effect of an anti $3$-brane space-time 
via an extra dimension is indeed compelling on a 3-brane universe. Analysis may reveal that a brane universe geometry appears to be influenced by an axion which is a potential candidate to source the dark energy.

%%%%%%%%%%%%%%%%%%%%%%%%%%%%%%%%%%%%%%%%%%%%%%%%%%%%%%%%%%%%%%
\subsection{Emergent gravity on a vacuum pair}
%%%%%%%%%%%%%%%%%%%%%%%%%%%%%%%%%%%%%%%%%%%%%%%%%%%%%%%%%%%%%%
The commutator $\left [ {\cal D}_{\mu}\ ,\ {\cal D}_{\nu}\right ]$ is worked out to yield a fourth order reducible torsion curvature tensor in a second order formalism \cite{abhishek-JHEP} underlying a vacuum created gravitational pair of $(3{\bar 3})$-brane \cite{abhishek-PRD}. With an appropriate re-scaling of ${1\over2}{\cal H}\rightarrow {\cal H}$, the emergent torsion curvature tensor takes a form: 
\be
{{\tilde{\cal K}}_{\mu\nu\lambda}{}}^{\rho}= \partial_{\mu}{{\cal H}_{\nu\lambda}}^{\rho} -\partial_{\nu} {{\cal H}_{\mu\lambda}}^{\rho} + {{\cal H}_{\mu\lambda}}^{\sigma}{{\cal H}_{\nu\sigma}}^{\rho}-{{\cal H}_{\nu\lambda}}^{\sigma}{{\cal H}_{\mu\sigma}}^{\rho}.\label{gauge-6}
\ee
The fourth order tensor is antisymmetric under an exchange of indices within a pair and is not symmetric under an exchange of its first pair of indices with the second. It differs from the Riemannian tensor $R_{\mu\nu\lambda\rho}$. However for a non-propagating torsion 
${\tilde {\cal K}}_{\mu\nu\lambda\rho}\rightarrow R_{\mu\nu\lambda\rho}$. The irreducible torsion curvature scalar ${\tilde K}$ is worked out on $S^1$ 
to describe a vacuum created gravitational pair of $(3{\bar 3})$-brane. Then, the local dynamics is approximated by two independent $4D$ curvatures 
and is given by
\be
S_{(3-{\bar 3})}= {1\over{12\kappa^2}}\int d^4x {\sqrt{-g}}\ \left ( {\cal K}\ -\ 3{\bar{\cal F}}_{\mu\nu}
{\cal F}^{\mu\nu} \right )\ ,\label{gauge-7}
\ee
where ${\cal F}_{\mu\nu}=\left ({\cal D}_{\mu}A_{\nu}-{\cal D}_{\nu}A_{\mu}\right )
= \left ( F_{\mu\nu} + {{\cal H}_{\mu\nu}}^{\alpha}A_{\alpha}\right )$. The $U(1)$ gauge invariance of the Lorentz scalar ${\cal H}^2$ has been shown to incorporate a metric fluctuation $f^q_{\mu\nu}= C(2\pi\alpha'){H_{\alpha\beta}}^{\mu} {{\cal H}^{\alpha\beta\nu}}$ to the background open string metric. It takes a form: 
\be
G_{\mu\nu}=\left ( G^{(NS)}_{\mu\nu} +\ C {\bar{\cal H}}_{\mu\lambda\rho}{{\cal H}^{\lambda\rho}}_{\nu}\right )\ ,\label{gauge-7}
\ee
where $B_{\mu\nu}^{(NS)}$ signifies a constant NS field and appropriately couples to an electromagnetic field $F_{\mu\nu}^{\rm linear}$ to retain the $U(1)$ gauge invariance. Thus $f^q_{\mu\nu}$ incorporates a non-perturbative correction to the open string metric established on a $D$-brane \cite{seiberg-witten}. Interestingly a low energy limit in the string-brane model corresponds to a non-propagating geometric torsion and hence the emergent scenario on a pair of $(3{\bar 3})$-brane is described with a large extra dimension. Then a $D$-brane or an anti $D$-brane correction, sourced by a geometric torsion, may be ignored to describe Einstein vacuum in the string-brane model. A fact that a torsion is dual to an axion on a gravitational $3$-brane ensures one degree of freedom. In addition ${\cal F}_{\mu\nu}$ describes a geometric one form field with two local degrees on a gravitational $3$-brane. A precise match among the (three) local degrees of torsion in ${\tilde{\cal K}}$ on $S^1$ with that in ${\cal K}$ and ${\cal F}_{\mu\nu}$ reassure the absence of a dynamical dilaton field in the emergent metric scenario under $S^1$ compactification. 

%%%%%%%%%%%%%%%%%%%%%%%%%
\section{Quintessential cosmology}
%%%%%%%%%%%%%%%%%%%%%%%%%
We revisit the domain of observational cosmology to reiterate a variable vacuum energy density in our universe. A time variation of cosmological vacuum energy density in the early epoch leading to a de Sitter geometry has been argued in the inflationary models \cite{quevedo-lecture,Shiu-tye-BraneInflation2,majumdar-davis-2,padmanabhan,cai}. Remarkably a hidden fifth essence, $i.e.$ a quintessence, is known to describe a time varying spatially homogeneous cosmic density with a negative pressure and a positive energy density. Quintessence is indeed a potential candidate to describe the dark energy in our universe \cite{quintessence,deobratijirset}.

\sp
\noindent
In the recent past a slowly rolling time dependent scalar field is believed to describe the quintessence presumably due to its simplest form as a tensor field which couples minimally to Einstein gravity \cite{chen-pan-jing,huan-jun-quintessence}. The acceleration of universe is known to be parameterized by the cosmological equation of state: $\omega=P/\rho$, where $P=-T^i_i<0$ and $\rho= T^0_0>0$ respectively define the pressure and energy density sourced by a scalar field. In the recent years the observational cosmology primarily focuses on the measurement of $\omega$ sourced by a quintessence. Its value is known to be bounded, $i.e.\ -1<\omega<-1/3$.

%%%%%%%%%%%%%%%%%%%%%%%%%%%%%%%%%%%%%%%%%%%%%%%%%%%
\subsection{Axionic scalar in a gravitational pair}
%%%%%%%%%%%%%%%%%%%%%%%%%%%%%%%%%%%%%%%%%%%%%%%%%%%%
Inflation suggests that axions were created abundantly during the Big Bang. Hence, axions could plausibly explain the dark matter problem of physical cosmology\cite{deobratijirset,duffy,sikivie}.
In the context, the quintessence energy density underlying a five dimensional geometric torsion on $S^1$ has been explored in a string-brane model \cite{abhishek-JHEP,abhishek-PRD,abhishek-NPB-P}. It was shown that the non-trivial torsion curvature scalar ${\cal K}$ has been shown to be sourced by a massless axion on an anti $D_3$-brane. In particular, the action (\ref{gauge-7}) underlying a vacuum created gravitational pair of $(3{\bar 3})$-brane reveals that the emergent $3$-brane can be governed by a geometric Lorentz scalar ${\cal F}^2$ while an anti $3$-brane can be described by the curvature scalar ${\cal K}$ or vice-versa. Then the two within the three local degrees of the five dimensional NS field would like to describe an emergent metric fluctuation on a $3$-brane. The remaining local degree underlying an axionic scalar on an anti $3$-brane becomes hidden to the emergent $(3+1)$-dimensional brane universe. Thus a dynamical axion on an anti-brane indeed influences our brane-world. In other words, an axion may naturally be identified with a quintessence scalar in a string-brane model which describes a vacuum created gravitational pair of $(3{\bar 3})$-brane.

\sp
\noindent
An extra fifth spatial dimension is transverse to the vacuum created gravitational pair of $(3{\bar 3})$-brane universe. It is known to be sourced by a higher dimensional gauge field and in the case it is a KR field on a $D_4$-brane in a string-brane setup. An axion is Poincare dual to a torsion on a gravitational anti $3$-brane and hence is inaccessible to an observer on a vacuum created $3$-brane. The idea of a quintessence energy density sourced by an axion field on an anti-brane is remarkable. The dynamical aspects of an axion on a $3$-brane universe may also be viewed via an extra fifth dimension in between the gravitational pair, which in turn is known to describe a Liouville scalar in string theory.

\sp
\noindent
A fact that the dark energy permeates all the space and tends to enhance the rate of expansion of universe may be understood via an extra dimension between a vacuum created pair of $(3{\bar 3})$-brane. Hubble's law may be revisited to explain the receding of a brane universe from an anti-brane. The law states that the recessional speeds of a brane and anti-brane universes are proportional to the transverse distance between them. Since a vacuum created brane universe moves away from its an anti-brane in an opposite direction, it implies a repulsive gravity between them. The notion of repulsive gravity has been argued to be sourced by a quintessence scalar field in cosmology. It provokes thought to believe that an axion in a hidden anti-brane is realized through a geometric torsion which in turn is responsible for the repulsive gravity. In fact a repulsive gravity underlying a quintessence has been conjectured between a matter and anti-matter which may seen to be in conformity with the notion of a vacuum created pair of $(D{\bar D})_3$-brane by a KR field quanta on a $D_4$-brane. Interestingly an effective Schwarzschild de Sitter black hole obtained in a gauge choice for a nonpropagating torsion in ref.\cite{priyabrat-EPJC} has been shown to describe a varying positive curvature scalar $R=12/r^4$. It further strengthen the presence of quintessence axion in a string-brane model. It may imply that the brane/anti-brane universe(s) at its vacuum creation, $i.e.$ at the past horizon radius $b$ with a Big Bang, was described by a constant curvature $R=12/b^4>0$. Computations \cite{priyabrat-IJMPA} have been performed to show that the quintessence kinetic energy can be ignored at the Big Bang and hence the equation of state reduces to $\omega=-1$. The result is in agreement with the value of $\omega$ known for a constant vacuum energy density \cite{chen-pan-jing,huan-jun-quintessence}.

\sp
\noindent
On the other hand a $D$-instanton is sourced by an interacting axionic scalar field $\psi$ on a gravitational anti $3$-brane within a vacuum created pair in the framework \cite{deobratijirset}. A space-like event on a $3$-brane universe becomes time-like across the event horizon where the pair was created by the KR form quanta. Thus the axion may appropriately be described solely by its time dependence. Interestingly the gravitational brane/anti-brane frame-work naturally incorporates the Friedman-Robertson-Walker (FRW) line-element and hence a scale factor $a(t)/a_0$. The FRW line-element, underlying the isotropic and homogeneous space-time, may be given by
\be
ds^2= -dt^2 + {a^2\over{a^2_0}} \left ( {{dr^2}\over{1-kr^2}} + r^2 d\Omega^2\right )\ ,\qquad {\rm where}\quad k={0,\pm 1}\ .\label{axion-0}
\ee
The axionic scalar dynamics, in presence of a background FRW metric, may be given by
\be
S_{\psi}= -{1\over2}\int d^4x\sqrt{-g}\ \left ( \left (\nabla\psi\right )^2 -V(\psi)\right )\ .\label{axion-1}
\ee
Apriori the energy-momentum-stress tensor $T_{\mu\nu}$ is worked out to yield:
\bea
T_{\mu\nu}&=&{{-2}\over{{\sqrt{-g}}}}\ {{\delta S_{\psi}}\over{\delta g^{\mu\nu}}}\nonumber\\
&=& \nabla_{\mu}\psi\nabla_{\nu}\psi - {1\over2}g_{\mu\nu}\left ( \left (\nabla\psi\right )^2 -V(\psi)\right )\ .\label{axion-2} 
\eea
Considering the system as an ideal gas, the thermodynamical equation of state for a constant temperature may be expressed in terms of axionic QFT dynamics in the background FRW metric. It may be recalled that a gravitational pair $(3{\bar 3})$-brane was vacuum created across a horizon where the light cone flips. Thus a space-like event on a gravitational $3$-brane becomes time-like on an anti $3$-brane which is precisely governed by a time dependent axionic scalar field $\psi(t)$. Then, the thermodynamic state parameter $\omega$ is given by
\bea
\omega&=&{{\rho}\over{p}}={{T_{00}}\over{T_{ij}}}\ ,\nonumber\\
&=&{{{\dot{\psi}}+ V(\psi)}\over{\dot{\psi}-V(\psi)}}\ ,\label{axion-3} 
\eea
The isotropic space, $i.e.$ FRW metric, enforces a normal stress $T_{ij}=Pg_{ij}$ and hence $T_i^j=Pg_i^j$. The momentum conservation ensures that an anti $3$-brane across a past horizon moves away from a $3$-brane universe which is in agreement with the repulsive quintessence. The fact may be perceived in a limit for a slow rolling axion. The limit ensures $\omega=-1$ and hence a vacuum energy dominated era. Nevertheless the dynamical axionic correction in the matter dominated universe plays a vital role and brings down the value of $\omega$ within the observed range $-1<\omega<-1/3$ and hence the quintessential cosmology becomes significant. Thus an axionic scalar interaction potential $V(\psi)$ plays a dominant role to ensure a negative pressure $P$ and hence the observed quintessence range for $\omega$. Interestingly the quantum correction, sourced by an axion, ensures the NS form dynamics on an emergent anti $3$-brane. A dynamical NS form incorporates a closed string coupling to a $D_3$-brane and has been identified as an emergent gravitational, or fat, $3$-brane. 

%%%%%%%%%%%%%%%%%%%%%%%%%%%%%%%%%%%%%%%%%%%%
\subsection{$D$-instanton ${\mathbf{\rightarrow}}$ Quintessence}
%%%%%%%%%%%%%%%%%%%%%%%%%%%%%%%%%%%%%%%%%%%%
An axion in the Ramond-Ramond (RR) sector of type II superstring ensures the significance of a D-instanton in the string-brane model. Importantly a D-instanton is known to incorporate a non-perturbative correction in a perturbative string theory. In fact the universe was argued to begin at a past horizon with a Big Bang on a pair of $(D{\bar D})$-instanton at an early epoch in de Sitter underlying a geometric torsion \cite{abhishek-JHEP}. The hot de Sitter is believed to expand and the notion of a real time was incorporated subsequently into the geometry. The expansion of a pair of brane (or anti-brane) universe from its origin at the Big Bang singularity was argued to pass through a series of nucleation of a pair of $D$-instanton followed by higher dimensional pairs such as $D$-particle, $D$-string and $D$-membrane. The nucleation process stops when a pair of $(3{\bar 3})$-brane is created as the pair with a transverse fifth dimension completely fills the $D_4$-brane world-volume.

\sp
\noindent
In other words an instanton local degree on an anti $D_3$-brane influences the world-volume of a $D_3$-brane within a pair. Arguably the instantaneous effect of a $D$-instanton on a $D_3$-brane is in agreement with the non-perturbative universe in $4D$. A quintessence axion, presumably on a vacuum created ${\bar 3}$-brane, is argued to source an extra fifth dimension to our brane universe on a $3$-brane with a vacuum pair. The curvature scalar ${\cal{K}}$ on a gravitational anti $3$-brane may describe a quintessence axion and a geometric ${\cal F}_2$ governs the torsion curvature on a gravitational $3$-brane. An accelerated and expanding universe in the present day cosmology may seen to be influenced by a growth in extra dimension between a brane and an anti-brane universes. The high energy torsion modes have been argued to decouple with a large fifth dimension and eventually a low energy brane universe may appear to decouple from the anti-brane in the limit. The decoupling of non-perturbative quantum effects may lead to describe Einstein vacuum in the setup.

\sp
\noindent
Furthermore the brane geometries have been shown to be sourced by an axionic scalar underlying a vacuum created gravitational pair of $(3{\bar 3})$-brane. The quantum effects have been worked out to show that the low energy perturbative string vacuum receives a non-perturbative $D$-brane world-volume correction in the string-brane model. Axionic sourced quantum geometries 
are further analyzed for two larger length scales to qualitatively describe a laboratory black hole in a medium energy regime and a semi-classical brane universe in a low energy regime. Analysis under the emergent metric scalings have been performed to explore the effective (Anti) de Sitter brane vacua in presence of non-perturbative correction(s). A quantum correction has been shown to be approximated by a varying energy density underlying an axionic quintessence. Thus an axion quanta can source a $D$-instanton which is the fundamental building block in type IIB superstring theory. The nucleation of a higher dimensional pair of brane/anti-brane ceases with a stable pair of $(3{\bar 3})$-brane underlying a five dimensional gauge theory.

%%%%%%%%%%%%%%%%%%%%%%%%%%%%
\section{Concluding remarks}
%%%%%%%%%%%%%%%%%%%%%%%%%%%%
We have briefly reviewed a geometric construction underlying a string-brane setup, formulated by us in collaborations, to address the quitessential cosmology. In particular the construction, underlying a modified covariant derivative, has been argied to incorporate a non-perturbative quantum correction to the low energy string vacuum in the recent past. Interestingly Schwinger pair production mechanism was invoked in the construction to describe an emergent gravitational pair of $(3{\bar 3})$-brane at the cosmological horizon by a KR field quanta in the world-volume gauge theory on a $D_4$-brane. A stable pair of brane and anti-brane universe has been known to be separated by extra transverse
dimension(s) in a ten dimensional superstring theory. They are believed describe the transverse scalars hidden to our brane-universe. One such hidden scalar in the frame-work has been identified with a dynamical axion on an anti $3$-brane within a pair. The axion was argued 
to describe the quintessence in the model which is believed to be a potential candidate for dark energy in universe. However the quintessence axion 
is known to source a $D$-instanton, which in turn incorporates non-perturbative correction to the matter dominated universe. At an early epoch, $i.e.$ right after a Big Bang, our universe is believed to be influenced significantly by a non-perturbative quantum correction sourced by an axion. Primarily the quantum corrections have been known to be described by a geometric torsion which forms a condensate in a low energy (GTR) limit.

\sp
\noindent
Generically the brane-world models have been successful to address the issues related to the cosmological inflation 
\cite{quevedo-lecture,Shiu-tye-BraneInflation2,majumdar-davis-2}. These models have been explored to describe the inflationary cosmology underlying a rapid accelerated expansion of our universe \cite{priyabrat-EPJC,priyabrat-IJMPA}. Interestingly a quintessence axionic scalar dynamics is hidden or outside the $4D$ brane universe in our frame-work. An emergent pair of gravitational brane and anti-brane universe are non-BPS and they may be viewed via a spontaneous supersymmetry breaking phenomenon as they possess their origin in a supersymmetric or BPS $D_4$-brane. Thus the  massless KR field regains mass, and renamed as a NS field, in an emergent supersymmetry broken phase underlying the gravitational dynamics. The massive NS field further reassures Newtonian gravity which is indeed a non-relativistic limit of the GTR. Intuitively, the dark energy sourced by boson(s) and the dark matter sourced by fermion(s) are intimately related by a plausible supersymmetry in a higher dimensional gauge theory. It was shown that a geometric torsion governs an effective space-time curvature scalar ${\cal K}$ on $S^1$. It may ensure an underlying $T$-duality symmetry between the gauge theory and the emergent gravity theory. Nevertheless our analysis leading to our proposition for an emergent, non-BPS, stable pair of gravitational $3$-brane and anti $3$-brane universe inspires thought and possibly sheds light to a non-perturbative formulation such as $M$-theory and needs further attention in the current literature. Our analysis signifying an emergent geometric torsion curvature appears to be an insightful tool to explore the cosmological domain related to the accelerated expansion of our universe. Generically a non-perturbative dynamical correction in the disguise of a $D$-instanton may provide a clue to unfold the higher or hidden (H) essence(s) as a guiding principle to arrive at the origin for dark energy in universe.

\sp
\noindent
Note: The data used to support the findings of this study are included within the article.

%**********************************************************%
\def\anp{Ann. of Phys.}
\def\cmp{Comm.Math.Phys.\ {}} {}
\def\springer{Springer.Proc.Phys.}
\def\prl{Phys.Rev.Lett.}
\def\prd#1{{Phys.Rev.} {\bf D#1}}
\def\jhep{JHEP\ {}}{}
\def\cqg{Class.\& Quant. Grav.}
\def\plb#1{{Phys. Lett.} {\bf B#1}}
\def\npb#1{{Nucl. Phys.} {\bf B#1}}
\def\mpl#1{{Mod. Phys. Lett} {\bf A#1}}
\def\ijmpa#1{{Int.J.Mod.Phys.} {\bf A#1}}
\def\ijmpd#1{{Int.J.Mod.Phys.} {\bf D#1}}
\def\mpla#1{{Mod.Phys.Lett.} {\bf A#1}}
\def\rmp#1{{Rev. Mod. Phys.} {\bf 68#1}}
\def\jaat{J.Astrophys.Aerosp.Technol.\ {}} {}
\def \epj#1{{Eur.Phys.J.} {\bf C#1}} 
\def \jcap{JCAP\ {}}{}
%**********************************************************%

\end{document}